\documentstyle[aps,multicol,psfig]{revtex}
\begin{document}
\draft

\title{Deviation from Maxwell Distribution in Granular Gases\\ with
  Constant Restitution Coefficient}

\author{ Nikolai V. Brilliantov$^{1,2}$ and Thorsten P\"oschel$^{1}$}
\address{$^1$Humboldt-Universit\"at zu Berlin, Institut f\"ur Physik,
  Invalidenstra{\ss}e 110, \\ D-10115 Berlin, Germany,
  http://summa.physik.hu-berlin.de/$\sim$thorsten/}
\address{$^2$Moscow State University, Physics Department, Moscow
  119899, Russia} \date{\today} \maketitle
\begin{abstract}
  We analyze the velocity distribution function of force-free granular
  gases in the regime of homogeneous cooling when deviations from the
  Maxwellian distribution may be accounted only by leading term in the
  Sonine polynomial expansion. These are quantified by the magnitude
  of the coefficient $a_2$ of the second term of the expansion.  In
  our study we go beyond the linear approximation for $a_2$ and
  observe that there are three different values (three roots) for this
  coefficient which correspond to a scaling solution to the Boltzmann
  equation. The stability analysis performed showed, however, that
  among these three roots only one corresponds to a stable scaling
  solution. This is very close to $a_2$, obtained in previous studies
  in a linear with respect to $a_2$ approximation.

\end{abstract}
\begin{multicols}{2}

The granular gases, i.e., rarefied systems composed of inelastically 
colliding particles have been of particular interest during the last 
decade (e.g.~\cite{Cluster,NEBO97,NoijeErnst:97,GoldshteinShapiro95}). 
Compared to gases of elastically colliding particles, the dissipation of 
energy at inelastic collisions leads to some novel phenomena in these 
systems. One can mention clustering (e.g.~\cite{Cluster}), 
formation of vortex patterns (e.g.\cite{NEBO97}), etc. Before clustering 
starts, the granular gas being initially homogeneous, keeps for some 
time its homogeneity, although its temperature permanently decreases. 
This regime is called the homogeneous cooling regime (HC). 

In the present study we address the properties of the velocity distribution 
of granular particles in the regime of HC, such as the deviation from the 
Maxwellian distribution and the stability of the distribution function. 
We assume that the restitution coefficient $\epsilon$ does not depend 
on the impact velocity, i.e. that $\epsilon={\rm const}$. The properties 
of the velocity distribution for the system with the impact-velocity 
dependent coefficient of restitution (e.g.~\cite{epsviav}) will be addressed 
elsewhere~\cite{BPtobepub}. 

It is well known that granular gases in the HC regime do not reveal
Maxwellian distribution
(e.g.\cite{NoijeErnst:97,GoldshteinShapiro95,BreyDuftyKimSantos,EsipovPoeschel:97}).
The high-velocity tail is overpopulated
\cite{NoijeErnst:97,EsipovPoeschel:97}, while the main part of the
distribution is described by the sum of the Maxwellian one and the
correction to it, written in terms of the Sonine polynomial expansion
(e.g.~\cite{NoijeErnst:97,GoldshteinShapiro95,BreyDuftyKimSantos}).
Usually only the leading, second term, in this expansion is taken into
account~\cite{NoijeErnst:97,GoldshteinShapiro95,BreyDuftyKimSantos},
moreover in previous studies~\cite{NoijeErnst:97,GoldshteinShapiro95}
only linear analysis with respect to the coefficient $a_2$, which
refers to this second term has been performed. Finding that $a_2$,
obtained within the linear approximation, is small, the authors of
Ref.\cite{GoldshteinShapiro95,NoijeErnst:97} conclude {\em a
  posteriori} that the the linear approximation is valid.

In our approach we also assume that one can restrict oneself to the
leading term in the Sonine polynomial expansion and ignore the other.
However we go beyond the linear approximation with respect to the
coefficient $a_2$ and perform complete analysis within this level of
the system description. We observed that there are three different
values of $a_2$ exist which correspond to the scaling solution of the
Boltzmann equation. The stability analysis for the velocity
distribution function shows, however, that only one value of $a_2$
corresponds to a physically acceptable stable scaling solution.  The
stable solution is close to the result previously obtained within the
linear analysis~\cite{NoijeErnst:97}.

To introduce notations and specify the problem we briefly sketch the
derivation of the coefficient
$a_2$~\cite{NoijeErnst:97,GoldshteinShapiro95}, in accordance with the
approach developed in Ref.\cite{NoijeErnst:97}.

So far we introduce the (time-dependent) temperature $T(t)$, and 
thermal velocity $v_0(t)$, which are related to the velocity 
distribution function $f({\bf v}, t)$ as 

\begin{equation}
\label{deftemp1}
\frac{3}{2} n T(t)=\int d {\bf v} \frac{v^2}{2} f({\bf v},t)
=\frac32 n v_0^2(t)
\end{equation}
here $n$ is the number density of the granular gas, the particles are
assumed to be of a unit mass ($m=1$), and (\ref{deftemp1}) is written
for 3D-systems. The inelasticity of collisions is characterized by the
coefficient of the normal restitution $\epsilon$ (here we consider
smooth particles), which relates after-collisional velocities ${\bf
  v}_1^*$, ${\bf v}_2^*$ to the pre-collisional ones, ${\bf v}_1$,
${\bf v}_2$ as:

\begin{equation}
\label{directcoll}
\label{v1v2v1*v2*}
{\bf v}_{1/2}^*={\bf v}_{1/2} \mp 
\frac12 (1+\epsilon) ({\bf v}_{12} \cdot {\bf e}){\bf e} 
\end{equation}
where ${\bf v}_{12}={\bf v}_1-{\bf v}_2$ is the relative velocity, the
unit vector ${\bf e} = {\bf r}_{12}/ |{\bf r}_{12}|$ gives the
direction of the vector ${\bf r}_{12}={\bf r}_{1}-{\bf r}_{2}$ at the
instant of the collision.  The time-evolution of the velocity
distribution function is subjected to the Enskog-Boltzmann equation,
which for the force-free case reads~\cite{NoijeErnst:97,resibua}:

\begin{eqnarray}
\label{collint} 
&&\frac{\partial}{\partial t}f({\bf v},t) =
g_2(\sigma)\sigma^2 \int d {\bf v}_2 \int d{\bf e}
\Theta(-{\bf v}_{12} \cdot {\bf e}) |{\bf v}_{12} \cdot 
{\bf e}| \nonumber \\
&& \times \, \left\{\frac{1}{\epsilon^2} f({\bf v}_1^{**},t)
f({\bf v}_2^{**},t)-
f({\bf v}_1,t)f({\bf v}_2,t) \right\} 
\end{eqnarray}
where $\sigma$ is the diameter of particles,
$g_2(\sigma)=(2-\eta)/2(1-\eta)^3$ ($\eta=\frac16\, \pi n \sigma^3$ is
packing fraction) denotes the contact value of the two-particle
correlation function~\cite{CarnahanStarling}, which accounts for the
increasing collision frequency due to the excluded volume effects;
$\Theta(x)$ is the Heaviside function. The velocities ${\bf v}_1^{**}$
and ${\bf v}_2^{**}$ refer to the precollisional velocities of the
so-called inverse collision, which results with ${\bf v}_1$ and ${\bf
  v}_2$ as the after-collisional velocities.  The factor
$1/\epsilon^2$ in the gain term appears respectively from the Jacobian
of the transformation $d{\bf v}_1^{**}d{\bf v}_2^{**} \to d{\bf v}_1
d{\bf v}_2$ and from the relation between the lengths of the
collisional cylinders $\epsilon |{\bf v}_{12}^{**} \cdot {\bf e}|
dt=|{\bf v}_{12} \cdot {\bf e}|dt$~\cite{NoijeErnst:97,resibua}.

Assuming that the velocity distribution function is of a scaling form:

\begin{equation}
\label{scal}
f({\bf v}, t)=\frac{n}{v_0^3(t)}\tilde{f}({\bf c})
\end{equation} 
one can show, that the scaling function satisfies the {\em
  time-independent} equation~\cite{NoijeErnst:97}:

\begin{equation}
\label{geneqveldis}
\frac{\mu_2}{3} 
\left(3 + c_1 \frac{\partial}{\partial c_1} \right) \tilde{f}({\bf c}) =
\tilde{I}\left( \tilde{f}, \tilde{f} \right)
\end{equation}
with the dimensionless collision integral:

\begin{eqnarray}
\label{dimlcolint}
&&\tilde{I}\left( \tilde{f}, \tilde{f} \right)=
\int d {\bf c}_2 \int d{\bf e}
\Theta(-{\bf c}_{12} \cdot {\bf e}) |{\bf c}_{12} \cdot {\bf e}| 
\nonumber \\
&& \times \, \left\{\epsilon^{-2} \tilde{f}({\bf c}_1^{**}) 
\tilde{f}({\bf c}_2^{**})-
\tilde{f}({\bf c}_1)\tilde{f}({\bf c}_2) \right\} 
\end{eqnarray}
and with the moments of the dimensionless collision integral
\cite{NoijeErnst:97}:

\begin{equation}
\label{mup}
\mu_p \equiv - \int d {\bf c}_1 c_1^p 
\tilde{I}\left( \tilde{f}, \tilde{f} \right)\ , 
\end{equation}
while the time-evolution of temperature reads:

\begin{equation}
\label{dTdt}
dT/dt=-(2/3)\, BT\mu_2
\end{equation} 
where $B=B(t) \equiv v_0(t) g_2(\sigma) \sigma^2 n$. 

To proceed we use the Sonine polynomial expansion for the velocity 
distribution function~\cite{NoijeErnst:97,GoldshteinShapiro95}

\begin{equation}
\label{genSoninexp}
\tilde{f}({\bf c})
=\phi(c) \left\{1 + \sum_{p=1}^{\infty} a_p S_p(c^2) \right\}
\end{equation}
where $\phi(c) \equiv \pi^{-d/2} \exp(-c^2)$ is the Maxwellian
distribution and the first few Sonine polynomials read: $S_0(x)=1$,
$S_1(x)=-x^2 +\frac32 $,
$S_2(x)=\frac{x^2}{2}-\frac{5x}{2}+\frac{15}{8}$, etc.  Multiplying
both sides of Eq.~(\ref{geneqveldis}) with $c_1^p$ and integrating over
$d {\bf c}_1$, we obtain~\cite{NoijeErnst:97}:

\begin{equation}
\label{momeq}
\frac{\mu_2}{3} p \left< c^p \right> = \mu_p
\end{equation}
where integration by parts has been performed and where we define
\begin{equation} 
\label{nukp}
\left< c^p \right>  \equiv \int c^p \tilde{f}({\bf c}, t)  d{\bf c} \, .
\end{equation}
The odd moments $\left< c^{2n+1} \right> $ are zero, while the even
ones, $\left< c^{2n} \right> $, may be expressed in terms of $a_k$ with
$0 \leq k \leq n$. Calculations show that $\left< c^2 \right> =
\frac32$, implying $a_1=0$, according to the definition of the
temperature (\ref{deftemp1}) (e.g.~\cite{NoijeErnst:97}), and that
$\left< c^4 \right> = \frac{15}{4}\left( 1 + a_2 \right)$.

Now we assume, that the dissipation is not large, so that the
deviation from the Maxwellian distribution may be accurately described
only by the second term in the expansion (\ref{genSoninexp}) with all
high-order terms with $p>2$ discarded. Then (\ref{momeq}) is an
equation for the coefficient $a_2$.  Using the above results for
$\left< c^2 \right>$ and $\left< c^4 \right>$ it is easy to show that
Eq.~(\ref{momeq}) converts for $p=2$ into identity, while for $p=4$ it
reads:

\begin{equation}
\label{eqa2}
5\mu_2 \left(1+a_2 \right) - \mu_4 = 0
\end{equation} 
The coefficients $\mu_p$ may be expressed in terms of $a_2$ due to the
definition (\ref{mup}) and the assumption $\tilde{f}= \phi (c) [
1+a_2S_2(c^2)]$.  Using the properties of the collision integral one
obtains for $\mu_p$~\cite{NoijeErnst:97}:

\begin{eqnarray}
\label{mupa2}
&&\mu_p=-\frac12 \int d{\bf c}_1\int d{\bf c}_2 \int d{\bf e} 
\Theta(-{\bf c}_{12} \cdot {\bf e}) 
|{\bf c}_{12} \cdot {\bf e}| \phi(c_1) \phi(c_2) 
\nonumber \\
&&\left\{1+a_2\left[S_2(c_1^2)+S_2(c_2^2) \right] + 
a_2^2\,S_2(c_1^2)S_2(c_2^2) \right\}
\Delta (c_1^p+c_2^p) \nonumber
\end{eqnarray}
where $\Delta \psi({\bf c}_i) \equiv \left[\psi({\bf c}_i^*)-\psi({\bf
    c}_i) \right]$ denotes change of some function $\psi( {\bf c}_i)$
in a direct collision.  Calculations, similar to that, described in
\cite{NoijeErnst:97}, yield the following result (some detail are
given in~\cite{BPtobepub}):

\begin{equation}
\label{mu2A}
\mu_2=\sqrt{2 \pi} (1-\epsilon^2)\left(1+\frac{3}{16}a_2+
\frac{9}{1024}a_2^2 \right)
\end{equation}
and 
\begin{equation}
\label{mu4A}
\mu_4= 4 \sqrt{2 \pi}\left\{ T_1+a_2T_2+a_2^2T_3 \right\}
\end{equation} 
with 
\begin{eqnarray}
\label{T123}
&&T_1=\frac14(1-\epsilon^2)\left( \frac92+\epsilon^2 \right) \\
&&T_2=\frac{3}{128}(1-\epsilon^2)(69+10\epsilon^2)+\frac12(1+\epsilon) 
\nonumber \\
&&T_3=\frac{1}{64}(1+\epsilon) +\frac{1}{8192}(1-\epsilon^2)
(9-30\epsilon^2) 
\nonumber 
\end{eqnarray}
The coefficients $\mu_2$ and $\mu_4$ were provided in Ref.
\cite{NoijeErnst:97} up to terms of the order of ${\cal O} (a_2)$. One
obtains the coefficient $a_2$ in the Sonine polynomial expansion in
this approximation by substituting (\ref{mu2A},\ref{mu4A}) into
(\ref{eqa2}) and discarding in Eqs.~(\ref{mu2A},\ref{mu4A}) all terms
of the order of ${\cal O} (a_2^2)$:
\begin{equation}
\label{a2lin}
a_2^{\rm NE}=
\frac{16(1-\epsilon)(1-2\epsilon^2)}{81 -17\epsilon +30\epsilon^2
(1-\epsilon)}
\end{equation} 

Calculations including the next order terms ${\cal O} (a_2^2)$ in the
coefficients $\mu_2$ and $\mu_4$ show that Eq.~(\ref{eqa2}) is a cubic
equation, which for physical values of $\epsilon$, $0 \le \epsilon \le
1$, has three different real roots, as it shown on Fig.
\ref{fig:scheme}.
\begin{minipage}{8.5cm}
\begin{figure}[htb]
\centerline{\psfig{figure=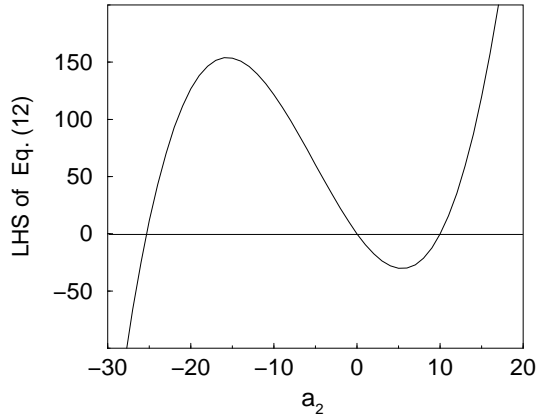,width=7cm}}
\caption{The left hand side of Eq.~\ref{eqa2} over $a_2$ for 
  $\epsilon=0.8$.  Obviously Eq.~\ref{eqa2} has three real solutions.}
\label{fig:scheme}
\end{figure}
\medskip
\end{minipage}

Although the cubic equation may be generally solved, the resultant
expressions for the roots are too cumbersome to be written explicitly.
However, one of the roots (the middle one) is rather small and close
to that given by Eq.~(\ref{a2lin}), obtained within the linear
approximation. This suggests the perturbative solution of the cubic
equation near this root:

\begin{equation}
\label{a2firord}
a_2=a_2^{\rm NE}\left[1-
\frac{1005(1-\epsilon^2)-4096T_3}{6080(1-\epsilon^2)-4096T_2}a_2^{\rm NE} 
+\cdots \right]
\end{equation} 
where we do not write explicitly terms of the order ${\cal
  O}\left([a_2^{\rm NE}]^3\right)$ and high-order terms. In
Fig.~\ref{fig:a2} the dependence of $a_2^{\rm NE}$ and of the
corresponding improved value $a_2$ are shown as a function of the
restitution coefficient $\epsilon$. As one can see from
Fig.~\ref{fig:a2} the maximal deviation between these is less than
$10\%$ at small $\epsilon$ and decreases as $\epsilon$ tends to $1$.

\begin{minipage}{8.5cm}
\begin{figure}[htb]
\centerline{\psfig{figure=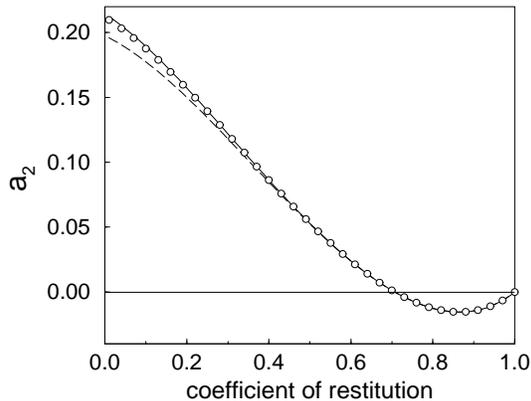,width=7cm}}
\caption{The second Sonine coefficient $a_2$ as a function of the 
  coefficient of restitution $\epsilon$ (full line). The dashed line
  shows $a_2^{\rm NE}$ in the first order approximation by van Noije
  and Ernst [3]
  according to Eq.~(\ref{a2lin}). The approximation (\ref{a2firord})
  is shown by circles.  }
\label{fig:a2}
\end{figure}
\medskip
\end{minipage}

The other two roots, shown on Fig.~\ref{fig:a2all} are of the order of
1 or 10, i.e. are not small. Physically, this means that one can not
cut the Sonine polynomial expansion in this case at the second term
and next order terms are not negligible to be discarded. Taking into
account the next order terms, i.e., releasing the assumption that $a_p
\simeq 0$ for $p>2$, breaks down the above analysis, since the
coefficients $\mu_2$, $\mu_4$ occur to be dependent not only on $a_2$,
but on $a_3$, $a_4, \ldots$ as well. Thus the occurrence of several
roots for the $a_2$, found within the above approach, which satisfy
the conditions required by the scaling ansatz (\ref{scal}) does not
imply the existence of several different scaling solutions.
Nevertheless such possibility may not be completely excluded. If one
assumes that few scaling distributions of the velocity may realize,
depending on the initial conditions at which the HC state has been
prepared, a natural question arises: Whether the particular scaling
solution is stable with respect to small perturbations, and what is
the domain of attraction of this particular scaling solution in some
parametric space.
\begin{minipage}{8.5cm}
\begin{figure}[htb]
\centerline{\psfig{figure=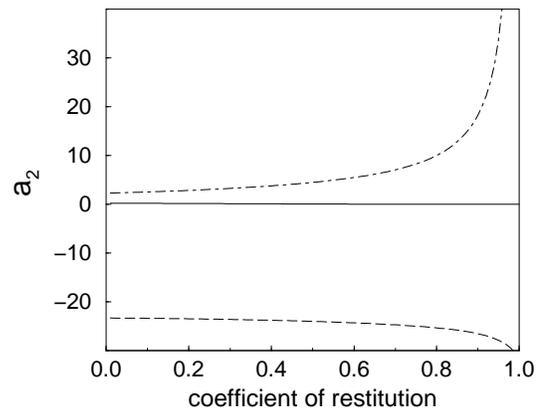,width=7cm}}
\caption{The other two solutions for second Sonine coefficient 
  $a_2$ of Eq.~\ref{eqa2} over the coefficient of restitution
  $\epsilon$. }
\label{fig:a2all}
\end{figure}
\medskip
\end{minipage}

Certainly, the stability problem is very complicated to be solved in
general. Therefore, we restrict ourselves to the stability analysis of
the scaling distribution (\ref{scal}) where the scaling function
$\tilde{f}({\bf c})$ has nonzero value of the coefficient $a_2$, while
the other coefficients $a_p$ with $p >2$ are negligibly small. (For
this scaling solution our above results for the coefficients $\mu_2$,
$\mu_4$ are valid).  Moreover, we assume, that small perturbations of
the (vanishingly small) coefficients $a_p$ with $p >2$ do not
influence the stability of the distribution, and analyze the stability
only with respect to variation of the coefficient $a_2$.

To analyze the stability of the velocity distribution we write it in a
more general form:

\begin{equation}
\label{genscal}
f({\bf v}, t)=\frac{n}{v_0^3(t)}\tilde{f}({\bf c}, t)
\end{equation} 
which leads, as it easy to show, to the following generalization of 
Eq.~(\ref{geneqveldis})~\cite{BPtobepub}:

\begin{equation}
\label{1geneqveldis}
\frac{\mu_2}{3} 
\left(3 + c_1 \frac{\partial}{\partial c_1} \right) \tilde{f}({\bf c}, t) +
B^{-1} \frac{\partial}{\partial t} \tilde{f}({\bf c}, t) =
\tilde{I}\left( \tilde{f}, \tilde{f} \right)
\end{equation}
with the collisional integral and coefficients $\mu_p$ being now
time-dependent.  The quantities $\left< c^p \right> $ also depend now
on time.  The temperature, however, evolves still according to
(\ref{dTdt}).

Using $\tilde{f}= \phi (c) [ 1+a_2(t)S_2(c^2)]$ and performing
essentially the same manipulations which led before to
Eq.~(\ref{eqa2}), we arrive at the following equation for the
coefficient $a_2(t)$:

\begin{equation}
\label{1eqa2}
\dot{a}_2-(4/3)\, B\mu_2 \left(1+a_2 \right)+(4/15) \, B\mu_4 =0
\end{equation} 
with $\mu_2$, $\mu_4$ still given by (\ref{mu2A},\ref{mu4A}), but with
the time-dependent coefficient $a_2(t)$. Writing the above value
$B(t)$ as
\begin{eqnarray} 
B(t)&=&(8\pi)^{-1/2}\tau_c(0)^{-1}u(t)^{1/2} \\
\tau_c(0)^{-1} &\equiv& 4 \pi^{1/2}g_2(\sigma)\sigma^2nT_0^{1/2}\,,
\end{eqnarray}
where $\tau_c(0)$ is related to the initial mean-collision time at the
initial temperature $T_0$, and $u(t) \equiv T(t)/T_0$ is the reduced
temperature, we recast Eq.~(\ref{1eqa2}) into the form:

\begin{equation}
\label{2eqa2}
\frac{da_2}{d \hat{t}}=\frac{\sqrt{2/ \pi}}{15}u^{1/2}F(a_2) 
\end{equation} 
where $\hat{t}$ is the reduced time, measured in units of $\tau_c(0)$,
and where we define a function:
\begin{equation}
\label{Fa}
F(a_2) \equiv  
5 \mu_2(1+a_2) -\mu_4\,.
\end{equation} 
The form of the function $F(a_2)$ for some particular value of
$\epsilon$ is shown on Fig.~\ref{fig:scheme}. This form of $F(a_2)$
persists for all physical values of the restitution coefficient, $0
\le \epsilon \le 1$.  This has three different roots,
$F(a_2^{(i)})=0$, $i=1,2,3$, which makes $d a_2/dt$ vanish yielding
the scaling form for the solution of the Enskog-Boltzmann equation.
The stability of the scaling solution, corresponding to $a_2^{(i)}$
requires for the derivative $dF/da_2$, taken at $a_2^{(i)}$ to be
negative, since only in this case a small deviation $a_2-a_2^{(i)}$
from $a_2^{(i)}$, corresponding to a scaling solution will decay with
time. As one can see from Fig.~\ref{fig:scheme} only the middle root,
which corresponds to small values of $a_2$, and is close to $a_2^{\rm
  NE}$, predicted by linear theory~\cite{NoijeErnst:97}, has negative
$dF/da_2$, and thus is stable. We also observed that for any $0 \le
\epsilon \le 1$ the point $a_2=0$ belongs to the attractive interval
of this stable root. Naturally, this means that initial Maxwellian
distribution will relax to the non-Maxwellian with $a_2 \approx
a_2^{\rm NE}$.

Note that relaxation of any (small) perturbation to this value of
$a_2$ occurs, as it follows from Eq.~(\ref{2eqa2}), on the collision
time-scale, i.e., practically ``immediately'' on the time-scale which
describes the evolution of the temperature. Therefore we conclude,
that the scaling solution of the Enskog-Boltzmann equation with $a_2$
corresponding to the middle root of the function $F(a_2)$, given with
a high accuracy by Eqs.~(\ref{a2firord},\ref{a2lin}), and with
negligibly small other coefficients $a_3$, $a_4, \ldots$ of the Sonine
polynomial expansion is a stable one with respect to (relatively)
small perturbations.

In conclusion, we analyzed the velocity distribution function of the
granular gas with a constant restitution coefficient at the regime of
the homogeneous cooling. We assume that the deviations from the
Maxwellian distribution may be described using only the leading term
in the Sonine polynomial expansion, with all other high-order terms
discarded. In this approach the deviations from the Maxwellian
distribution are completely characterized by the magnitude of the
coefficient $a_2$ of the leading term. We go beyond previous linear
theories and perform a complete analysis (on the level of the
description chosen), without discarding any nonlinear with respect to
$a_2$ terms.

Performing the stability analysis of the scaling solution of the
Enskog-Boltzmann equation we observe, that only one value of $a_2$,
obtained within our nonlinear analysis corresponds to a stable scaling
solution. We also report a corrections for this value of $a_2$ with
respect to the previous result of the linear theory. This corrections
are small (less than $10\%$) for all values of the restitution
coefficient $\epsilon$ and vanishes as $\epsilon$ tends to unity in
the elastic limit.

\end{multicols}
\end{document}